**Tunable Narrowband Terahertz Radiation from van der Waals Ferroelectrics**

Chun-Ying Huang[1,*], Taketo Handa[1,*], Daniel G. Chica[1,*], Zhihao Cui[1], Ding Xu[1], Jeongheon Choe[1], Yiliu Li[1], Margalit L. Feuer[1], Milan E. Delor[1], Michael Fechner[2], David R. Reichman[1], Xavier Roy[1,†], & Xiaoyang Zhu[1,†]

[1] Department of Chemistry, Columbia University, New York, NY 10027, USA
[2] Max Planck Institute for the Structure and Dynamics of Matter, Hamburg, Germany

[*] These authors contribute equally
[†] To whom correspondence should be addressed. emails: Xavier Roy: xr2114@columbia.edu;
 Xiaoyang Zhu: xyzhu@columbia.edu.

**Abstract:**

The terahertz (THz) spectral range is central to high-speed communication, precision metrology, sensing technologies, and a range of fundamental scientific investigations. Achieving these capabilities in practical systems increasingly demands chip-scale integration of THz photonic components that are typically bulky. In this context, van der Waals (vdW) materials provide a unique platform for integrated nonlinear photonics in the visible and near-infrared regimes, and extending this framework into the THz domain would constitute a significant advance. Here, we report tunable, intense, and narrowband THz radiation from ferroelectric niobium oxyhalides. Through halogen substitution and alloying, we achieve continuous and precise control over the emission frequency from 3.1 to 5.8 THz. We show that the narrowband THz radiation is driven by phonons associated with the ferroelectric polarization. We further demonstrate dynamic and nonvolatile control of the polarity of the coherent THz wave with external electric field. This work demonstrates efficient narrowband THz emission from vdW ferroeletrics and provides microscopic insight into its origin, paving the way for on-chip THz technology for a broad range of applications.

**Introduction:**

Van der Waals (vdW) materials have emerged as an attractive platform for nonlinear optics and photonics owing to their large optical nonlinearity, dynamical tunability, and ultracompact footprint.[1–4] Nonlinear optical properties of vdW materials, including transition metal dichalcogenides[4] and layered oxyhalides,[5,6] in the visible and near-infrared (NIR) regions have been extensively explored. There are compelling reasons to extend vdW photonics to the terahertz (THz) frequency regime. THz radiations are vital to high-speed communications, noninvasive diagnostics, sensing, metrology,[7] and fundamental sciences.[8–10] Generation of coherent THz radiation has been traditionally achieved with millimeter-sized nonlinear crystals, but recent works have demonstrated that ultrathin emitters can emit broadband THz radiation with high efficiencies.[11,12] Developing vdW THz emitting materials could enable the miniaturization of typically bulky THz optical components for chip integration and extreme THz light confinement.[13–15]

The vdW ferroelectric semiconductor $NbOI_2$ has recently emerged as a promising THz e mitter, with experimental observation of efficient and broadband THz emission,[16,17] and theoretical work predicting extraordinary electro-optic properties.[18] Handa et al. [16] demonstrated on-chip integration of thin $NbOI_2$ for *in situ* THz spectroscopy of a target vdW heterostructure, a step toward on-chip THz technology and its application to quantum science research, while complementing other integrated THz technologies.[19–21] A further intriguing feature of $NbOI_2$ is that it exhibits sharp THz emission at 3.13 THz with a width of less than 0.1 THz at room temperature for thicknesses smaller than several hundred nanometers.[16] This emission is explained as originating from coherent ferrons,[22] collective excitations of the ferroelectric order, suggesting a new paradigm of *ferronics*.[23–25] The narrowness of the emission is striking given that the thermal energy at room temperature is equivalent to ~6 THz, which would ordinarily produce substantial thermal broadening in electronic phenomena. These results suggest $NbOI_2$ may serve as an integrable, narrowband THz source that operates without the need for THz bandpass filters or phase-matching techniques,[26] opening new opportunities for integrated platforms for precision THz metrology,[7] spectral bioimaging,[27] and coherent driving of quantum phenomena.[9,28] Although these studies[16,22] point to a close connection between narrowband THz radiation and ferroelectric phonon modes in $NbOI_2$, the microscopic mechanism remains unclear and tunability of the narrowband emission has not been demonstrated.

Here, we report precise tuning of the narrowband THz emission in the family of vdW ferroelectric materials NbOX$_2$ (X = I, Br, Cl), elucidate the microscopic mechanism underpinning the emission, and demonstrate nonvolatile electrical control of THz signal. Across all compositions, we observe narrowband THz emission, with peak frequencies ranging from 3.13 to 5.85 THz. We combine first-principles calculation, analytical theory, and spectroscopic measurement to show that IR-active phonon modes with non-zero projection of transition dipole moments along the spontaneous polarization direction contribute to the narrowband radiation via a phonon-driven mechanism. By employing a halogen alloying approach,[29] we achieve precise tuning of the emission frequency. Moreover, we demonstrate electrical switching of the polarity of coherent THz radiation and confirm its nonvolatile memory behavior. These results establish vdW ferroelectrics as promising THz materials for integrated THz technologies and provide mechanistic insights and design strategies for THz light sources based on vdW materials and heterostructures.

**Results:**

Observation of multiple narrowband THz emission peaks

We synthesize millimeter-size single crystals of niobium oxyhalides NbOX$_2$ using the chemical vapor transport (CVT) method and characterize their structures by X-ray diffraction (XRD) (Supplementary Fig 1). The NbOX$_2$ family adopts a layered vdW structure with a monoclinic $C2$ space group. Fig. 1a shows the in-plane structure of NbOX$_2$, with Nb atoms off-center along the crystallographic $b$-axis, resulting in robust spontaneous polarization ($P_0$) at room temperature [30] and associated giant second-order nonlinear susceptibility ($\chi^{(2)}$).[5] The in-plane polarization $P_0$ of each layer adopts the same orientation across the stack. Consequently, NbOX$_2$ is a rare room-temperature vdW ferroelectric semiconductor with in-plane polarization, in contrast to most other materials, [31,32] which exhibit out-of-plane or mixed polarization.

Fig. 1b is a schematic for the THz emission experiments. We pump an exfoliated flake of NbOX$_2$ with a below-gap near infrared (NIR) laser centered at 800 nm with a pulse duration of 35 fs. The emitted THz radiation is collected, collimated, sent through a pair of wire-grid polarizers, and re-focused on a THz detector (Methods). Unless otherwise mentioned, the THz signals are detected by electro-optical (EO) sampling in a 200-μm-thick GaP detector. The NIR pump induces nonlinear polarization in NbOX$_2$, resulting in broadband THz radiation via optical rectification,[33] while also launching coherent ferrons [22] via an impulsive Raman mechanism.[34] We focus on mode(s) that modulate the spontaneous polarization ($P_0$) of the ferroelectric order (Fig. 1b).

Upon NIR excitation, we observe coherent THz radiation from NbOX$_2$ flakes with typical thicknesses of ~300 nm. Fig. 1c-e show time-domain waveform of emitted radiation parallel to the polar *b*-axis (red curves) from NbOI$_2$, NbOBr$_2$, and NbOCl$_2$, respectively. Following the initial optical rectification response (< 0.6 ps), we observe a pronounced and persistent oscillatory signal throughout the sampling window, set by the arrival of the echo signal at 4.5 ps. The corresponding frequency-domain spectra of the oscillating signals are shown in Fig. 1f-1h, obtained by Fourier transforming the time-domain data after 0.7 ps. Frequency-domain spectra over the entire sampling time are shown in Supplementary Fig. 2, where broadband emission due to optical rectification extends up to 8 THz, limited only by the detector bandwidth.[35] Handa et al.[16] established that the thickness normalized THz emission efficiency from optical rectification in NbOI$_2$ is 20-100x larger than that from the standard THz emitter ZnTe. The peak intensity of the narrow band THz emission is as much as 10x larger than that of the broad emission[16], pointing to exceptional efficiency for the former.

For NbOI$_2$ (Fig. 1f), the very narrow emission appears at 3.13 THz, which is the same as detected by ZnTe as an EO detector[22] and attributed to the collective excitation of ferroelectric order (detailed discussion is provided below).[20] Note that the small thickness of GaP EO detector and the associated short sampling-time window limits the current frequency resolution in Fig. 1f-h, while the actual linewidth is much narrower. As detailed in the next section, density functional theory (DFT) calculations establish a transverse optical (TO) phonon mode (Fig. 1b) characterized by a pronounced displacement of Nb along the *b*-axis and the associated modulation of the spontaneous polarization $P_0$. We label this ferroelectric TO phonon mode of NbOI$_2$ as F$_I$. Moreover, using GaP, which has a broader detection bandwidth than ZnTe, we identify an additional weak feature at 4.30 THz, denoted as P$_I$.

Similar to NbOI$_2$, NbOBr$_2$ and NbOCl$_2$ emit pronounced narrowband radiation at 3.68 THz (F$_{Br}$ in Fig. 1g) and 4.88 THz (F$_{Cl}$ in Fig. 1h), respectively. In the next section, we confirm these peaks correspond to the ferroelectric TO phonon modes. Substituting heavy iodine with lighter halogens (Br, Cl) results in reduced effective mass of the normal modes, leading to a blue-shift in the eigenfrequency and associated THz emission frequency,[36] while preserving the robust ferroelectricity. This is consistent with the eigenmode analysis in Fig. 1b, showing the involvements of halogen motions in the normal modes. These results imply that chemical modification can tune both the THz frequencies and ferroelectric properties of these materials, since the spontaneous polarization is governed predominantly by the Nb–O framework—consistent with our recent work showing the absence of ferroelectricity in $TaOBr_2$.[22]

Similar to $NbOI_2$, we also identify an additional emission peak for $NbOBr_2$ at 5.88 THz (labelled as $P_{Br}$), at higher frequency than the main peak $F_{Br}$ at 3.68 THz. The absence of a second emission peak in $NbOCl_2$ can be attributed to the limited detection range of our experiment beyond 8 THz.[26] Note that additional peaks are discernable at the low-frequency side of the main peak for all compounds: 1.78 THz for $NbOI_2$ (Fig. 1f); 2.45 THz for $NbOBr_2$ (Fig. 1g); and 3.92 THz for $NbOCl_2$ (Fig. 1h) (see also Supplementary Fig. 2 for higher frequency resolution). These features result from the re-absorption of broadband emission due to the TO phonon modes, according to the derivative spectral structures in the full-time window spectra (see Supplementary Fig. 2). We stress that the THz emission discussed above is along the polar axis. The blue curves in Fig. 1c-1h demonstrate this point: no THz emission is detected along the non-polar $c$-axis, apart from residual signals attributed to the finite extinction ratio of the polarizers (Methods).

Origin of narrowband radiation by THz absorption, Raman, and DFT calculations.

To elucidate the origin of the narrowband, polarized THz emission from $NbOX_2$, we turn to THz absorption spectroscopy, Raman spectroscopy, and first principles calculations. Fig. 2 shows calculated IR activities (a-c), experimental polarization-resolved IR absorption spectra (d-f), and unpolarized Raman spectra (g-i) of $NbOX_2$. The IR spectra are measured by THz time-domain spectroscopy (see Supplementary Fig. 3 for raw time-domain traces). The red and blue curves in Fig. 2a-f correspond to IR spectra along the polar $b$-axis and non-polar $c$-axis, respectively (see Methods for the symmetry assignment of the calculated spectra).

In the experimental THz absorption spectra collected with polarization along the polar $b$-axis (red curves in Figs. 2d-f), we identify strong and sharp absorptions at 3.13 THz for $NbOI_2$, 3.71 THz for $NbOBr_2$, and 4.85 THz for $NbOCl_2$. These frequencies match the main THz peaks in Fig. 1f-h. Note that THz radiation corresponds to the far-IR spectral region; for simplicity, we call this "IR" in the following discussions. These absorption features are also observed in the calculated IR activities in Fig. 2a-c. Comparing the calculated and experimental absorption spectra (Figs. 2a-c and 2d-f), we assign the main modes $F_I$, $F_{Br}$ and $F_{Cl}$, to those with the highest calculated IR activity in Fig. 2a-c. We identify weaker IR absorption features from both calculations and experiments at 4.30 and 5.88 THz for $NbOI_2$ (Fig. 2d) and $NbOBr_2$ (Fig. 2e), respectively. These frequencies match the $P_I$ and $P_{Br}$ peaks observed in the THz emission spectra in Figs. 1f and 1g. The oscillator strengths of $P_I$ and $P_{Br}$ in the absorption spectra are much smaller than those of $F_I$ and $F_{Br}$, consistent with the calculated IR activities.

Because all $NbOX_2$ crytals lack inversion symmetry, their optical phonons are both IR and Raman active. Consequently, the IR-active modes can be driven via an impulsive Raman excitation mechanism with a femtosecond laser pump. To confirm this scenario, we measure the Raman spectra (unpolarized) of $NbOI_2$, $NbOBr_2$, and $NbOCl_2$ (Figs. 2g-i), and identify the peaks corresponding to $F_I$, $F_{Br}$ and $F_{Cl}$, as well as $P_I$ and $P_{Br}$, at the same frequencies observed as those in the IR spectra. Note that the Raman intensities of $P_I$ and $P_{Br}$ are more than one order of magnitude weaker than those of $F_I$ and $F_{Br}$, and thus may not be resolved at low laser excitation intensities.[37] By performing polarization-angle-resolved Raman measurements (Supplementary Fig. 5) and phonon symmetry analysis (Supplementary Fig. 6), we confirm that $F_I$, $F_{Br}$, $F_{Cl}$, $P_I$, and $P_{Br}$ all belong to the *A* symmetry (see Methods for irreducible representations). As a result, these modes show IR activities along the polar *b*-axis,[38] in agreement with the calculated IR activities (Figs. 2a-c) and experimental IR absorption spectra (Figs. 2e-g). The full width at half maximum (FWHM) of the Raman ferroelectric mode decreases with X, from 0.095 THz for $NbOCl_2$ ($F_{Cl}$) to 0.039 THz for $NbOI_2$ ($F_I$). The large FWHM for $NbOCl_2$ indicates a shorter phonon lifetime than $NbOI_2$, consistent with the faster temporal decay of the THz emission in Fig. 1e. In contrast, we observe negligible decay of the THz emission within the sampling time and an ultra-sharp Raman feature for $NbOI_2$, pointing to an unusually *harmonic* nature of this ferroelectric TO mode.

The behavior along the non-polar *c*-axis differs markedly for all $NbOX_2$. Although strong IR absorption peaks are observed (blue curves, Fig. 2d-f), we measure no THz emission at the corresponding frequencies (blue curves, Fig. 1f-h). The phonon symmetry analysis in Supplementary Figs. 5,6 confirms that the IR modes along the *c*-axis possess *B* symmetry.[38] These results show that IR-active optical phonons with *A* symmetry emit long-lived, coherent THz radiation along the spontaneous polarization, whereas IR-active modes with *B* symmetry do not.

Theoretical analysis.

To understand the microscopic mechanism, we analyze eigenvectors of the radiative and selected non-radiative phonon modes at the zone center. Fig. 3 shows the results for NbOI$_2$ and NbOBr$_2$, and Supplementary Fig. 7 shows those for NbOCl$_2$. For F$_I$ and F$_{Br}$ (Fig. 3a,d, respectively), we identify Nb and O atoms oscillating nearly parallel to the polar *b*-axis with significant contributions from the Nb atoms. Thus, these modes strongly modulate spontaneous polarization. For P$_I$ and P$_{Br}$ (Fig. 3b,e, respectively), the eigenvectors of these modes oscillate largely along the non-polar axis but are noticeably tilted. The Nb displacement vectors appreciably deviate from the non-polar direction (~ 23° for P$_{Br}$), while the O displacement vectors show only minor deviation (~ 7° for P$_{Br}$). Combined with the larger displacement amplitude of the Nb atoms, we find that P$_I$ and P$_{Br}$ modes result in oscillating dipoles with a finite projection along the polar axis.

In contrast, the eigenvectors for the *B*-symmetry modes at 5.35 and 6.23 THz for NbOI$_2$ NbOBr$_2$, respectively, oscillate primarily along the non-polar axis, with negligible projected oscillation amplitude along the polar axis (Fig. 3c,f). This is in agreement with the absence of IR oscillator strength of these modes along the *b*-axis and with the symmetry analysis. Modes that have little effect on the spontaneous polarization along the *b*-axis do not emit THz radiation.

To establish a microscopic picture, we analyze THz radiation based on a phononic framework.[39,40] We first write down the potential for an IR-active mode in a centrosymmetric system:

$$V = \frac{1}{2}\omega_{IR}^2 Q_{IR}^2 + (Z^* Q_{IR} + \alpha Q_{IR}^3)E + \frac{1}{2}\left(\chi_{ee}^{(1)} + \beta Q_{IR}^2\right)E^2, \qquad (1),$$

where $Q_{IR}$ is the IR mode coordinate, $Z^*$ is the mode effective charge, $\alpha$ describes the modulation of $Z^*$, $E$ is the excitation field, $\chi_{ee}^{(1)}$ is the electric (non-phonon) susceptibility, and $\beta$ characterizes its modulation with respect to a polar distortion. We then rewrite Eq. 1 with $P = Z^* Q_{IR}$, the resulting potential in terms of $P$ is:

$$V = \frac{1}{2}\omega_{IR}^2 \frac{P^2}{Z^{*2}} + \left(P + \alpha \frac{P^3}{Z^{*3}}\right)E + \frac{1}{2}\left(\chi_{ee}^{(1)} + \beta \frac{P^2}{Z^{*2}}\right)E^2, \qquad (2)$$

The resulting polarization is:

$$P = \frac{\partial V}{\partial E} = \left(P + \alpha \frac{P^3}{Z^{*3}}\right) + \left(\chi_{ee}^{(1)} + \beta \frac{P^2}{Z^{*2}}\right)E. \tag{3}$$

Next, we consider a ferroelectric TO mode and replace $P$ with $P_0 + \delta P$, where $P_0 = Z^* Q_{IR,0}$ represents the spontaneous polarization and $\delta P = Z^* \delta Q_0$ represents the excitation. The polarization now can be represented as:

$$P = P_0 + \frac{\alpha P_0^3}{(Z^*)^3} + \left(1 + \frac{3\alpha P_0^2}{(Z^*)^3}\right)\delta P + \cdots, \tag{4}$$

Note that we can neglect the linear terms in $E$ here as the frequency of the light pulses is out of resonance with the eigenfrequency of the TO mode. Moreover, we truncate the series to the linear term to represent the excitation of polarization. Next, we can address the source term for the coherent THz emission resulting from the temporal change of the polarization:

$$\frac{\partial^2 P}{\partial t^2} = \left(1 + \frac{3\alpha P_0^2}{(Z^*)^3}\right)(\delta P)'', \tag{5}$$

which indicates that the additional nonlinear contribution arises from the spontaneous polarization $P_0$. The relative magnitude of this term can be linked to the change in LO–TO splitting across the ferroelectric transition, which may increase by a factor of 5 to –10, and thus lead to a substantial enhancement of the emission.

We attribute the prominent, narrowband THz emission at the ferroelectric phonon modes ($F_I$, $F_{Br}$, $F_{Cl}$) to this additional ferroelectric contribution in Eq. (4). For the tilted modes with finite projection onto the polarization axis ($P_I$ and $P_{Br}$), the mode excitation still can have finite values $\delta P$, leading to narrowband THz emission. In contrast, the *B*-symmetry modes (Fig. 3c,f) have no IR activity along the polar axis and thus zero interaction with $Q_{IR,0}$. These orthogonal dynamics lead to no THz emission for these IR-active *B*-symmetry phonons. We additionally analyze in Supplementary Fig. 7 the lower-lying *A*-symmetry modes (1.78, 2.45, and 3.92 THz for NbOI$_2$ in Fig. 2d, NbOBr$_2$ in Fig. 2e, and NbOCl2 in Fig. 2f, respectively). These modes are dominated by halogen motion, with negligible involvement from the Nb and O atoms. The absence of emission from the lower-lying *A*-symmetry modes across the NbOX$_2$ family can be understood as follows: the lower ionicity of halogens compared with Nb and O produce smaller changes in the dipole moment, and their antiparallel displacements further suppress the net polarization induced by these modes. As a consequence, these *A*-symmetry modes do not emit appreciable THz radiation.

We can extend this model to other emissive IR and Raman phonons of the parent (non-ferroelectric) phase. The IR activities of $P_I$ and $P_{Br}$ arise only after the ferroelectric transition, which lower the symmetry through spontaneous polarization along the Nb–O axis. In this situation we can define an induced mode effective charge $Z^*_{ind} = P_0 R^*$, where $R^*$ is the effective Raman charge of the corresponding parent-phase Raman mode, which is typically small. The resulting induced polarization is then $P_{0,ind} = Z^*_{ind} Q_{Raman, ind}$, where $Q_{Raman, ind}$ is an induced Raman mode coordinate. Within this framework, any mode that acquires a distortion along the polar axis in the ferroelectric phase gains an effective dipole moment, and can therefore produce narrow-band emission when excited.

It is intriguing to note that only vdW ferroelectric materials ($NbOX_2$ in the present study and $WO_2X_2$[22,29]) show clear narrowband emission associated with ferroelectric phonon modes. One primary reason may be that the large thickness of the three-dimensional crystals used in previous studies results in strong reabsorption at the same TO phonon frequency, and the significant phase mismatch can mask the narrowband emission. The ferroelectric $LiNbO_3$ and $LiTaO_3$ have been extensively used as THz emitters, but no study shows narrowband emission resulting directly from their $A_1$ soft modes at 7.5 and 6.2 THz, respectively.[41] In Cherenkov-type radiation, THz emission from $LiNbO_3$ has been reported to be limited to frequencies below 7.2 THz, and emission from $LiTaO_3$ to below 6 THz.[42–44] An earlier study identified an oscillating waveform with a frequency of 4 THz from a 1-mm-thick $LiTaO_3$ crystal,[41] where the authors attributed this coherent waveform to the damping of nonlinear polarization resulting from an excitation of the soft TO phonon at 6.2 THz, and explained the frequency discrepancy by considering a spectral filtering by lattice absorption. Recent work on thin $LiNbO_3$ (~500 nm) samples showed THz emission but no sharp emission peak attributable to a TO mode.[45] Further theory and experimental works are necessary clarify these potential contributions.

Chemical tunability and nonvolatile electrical control of narrowband THz emission.

The large variation in narrowband THz emission from $NbOI_2$ to $NbOCl_2$ underscores the potential of niobium oxyhalides for continuous tuning across the THz range. To explore this possibility, we synthesized mixed-halogen $NbOBrI$ and $NbOClBr$ single crystals using chemical vapor transport (see Supplementary Methods). XRD confirms that both materials retain the crystal structure of the parent compounds (Supplementary Fig. 8). Energy-dispersive X-ray spectroscopy reveals Br:I and Br:Cl atomic ratios of 1.32 ± 0.20 and 1.10 ± 0.02 for NbOBrI and NbOClBr, respectively, and verifies compositional uniformity (Supplementary Figs. 9, 10; Supplementary Tables 1, 2).

Figs. 4a and 4b show the THz emission spectra of NbOBrI and NbOClBr together with the normalized emission spectra of their parent compounds. As shown in Fig. 4a, the emission frequency of NbOBrI lies near the midpoint between those of $NbOBr_2$ and $NbOI_2$. Likewise, NbOClBr exhibits THz emission intermediate between its two parent compounds (Fig. 4b). The emission from NbOClBr is broader than that of the other materials (see Supplementary Fig. 11 for time-domain traces of NbOClBr and NbOBrI), which we attribute to a relatively high degree of disorder in this alloyed compound. Overall, by combining parent and alloyed crystals of the vdW ferroelectric $NbOX_2$, we demonstrate tunable, narrowband THz emission over a continuous range between 3.1 and 5.8 THz.

As expected for a ferroelectric mechanism, the THz emission is controllable with an applied electric field. To demonstrate this, we fabricate a device consisting of a 120-nm-thick $NbOI_2$ flake contacted by metallic electrodes separated by 70 μm.[46] We evaluate ferroelectric switching behavior using electric field-dependent second harmonic generation (SHG) (Fig. 5b; Methods and Supplementary Fig. 12). By sweeping the electric field, we observe a clear hysteresis in the SHG intensity. At fields of +40 kV/cm and –20 kV/cm (i.e. the coercive fields),[43] the spontaneous polarization increases and decreases markedly, respectively, consistent with a previous report.[47] The asymmetry may be attributed to internal bias field resulting from strain between the electrodes and the flake, and between the flake and the substrate.[48,49] Near the coercive fields, fluctuation of the SHG intensity (seen in the spread of the SHG intensity in Fig. 5) is likely due to the inherent dynamic process of the structural change resulting from ferroelectric domain switching (see also Supplementary Fig. 13).

To demonstrate nonvolatile electric-field control of the narrowband THz radiation, we first record THz emission without an applied field using a 1-mm-thick ZnTe crystal as an EO detector (red curves in Fig. 5c,d for time- and frequency-domain spectra, respectively). We then apply –50 kV/cm to flip the ferroelectric polarization, return the device to zero field, and measure the THz emission with no applied field. This polarization switch flips the polarity of the THz radiation (Fig. 5c and Supplemental Fig. 14), and decreases the amplitude. We attribute the reduced amplitude to incomplete domain switching, which leads to partial cancellation of the THz signal.

Discussion and outlook

We have synthesized a family of vdW ferroelectric $NbOX_2$ crystals and investigated, both experimentally and theoretically, their narrowband THz radiation. We demonstrate chemically tunable narrowband THz emitters with a broad tuning range achieved through halogen substitution and alloying. Our spectroscopic measurements and first-principles calculations show that IR-active phonons with dipole oscillations aligned with the spontaneous polarization are responsible for the observed narrowband emission at room temperature. Our theoretical analysis further points to a nonlinear phononic origin for the emission, suggesting additional routes, such as strain engineering and chalcogenide substitution, for tuning the THz frequency. These findings open a pathway toward integrated THz technologies based on vdW materials for high-speed information transmission and fundamental quantum science.

*Note added*. We note that during the preparation of this manuscript, a preprint concerning THz emission properties from $NbOX_2$ was posted on arXiv.[50]

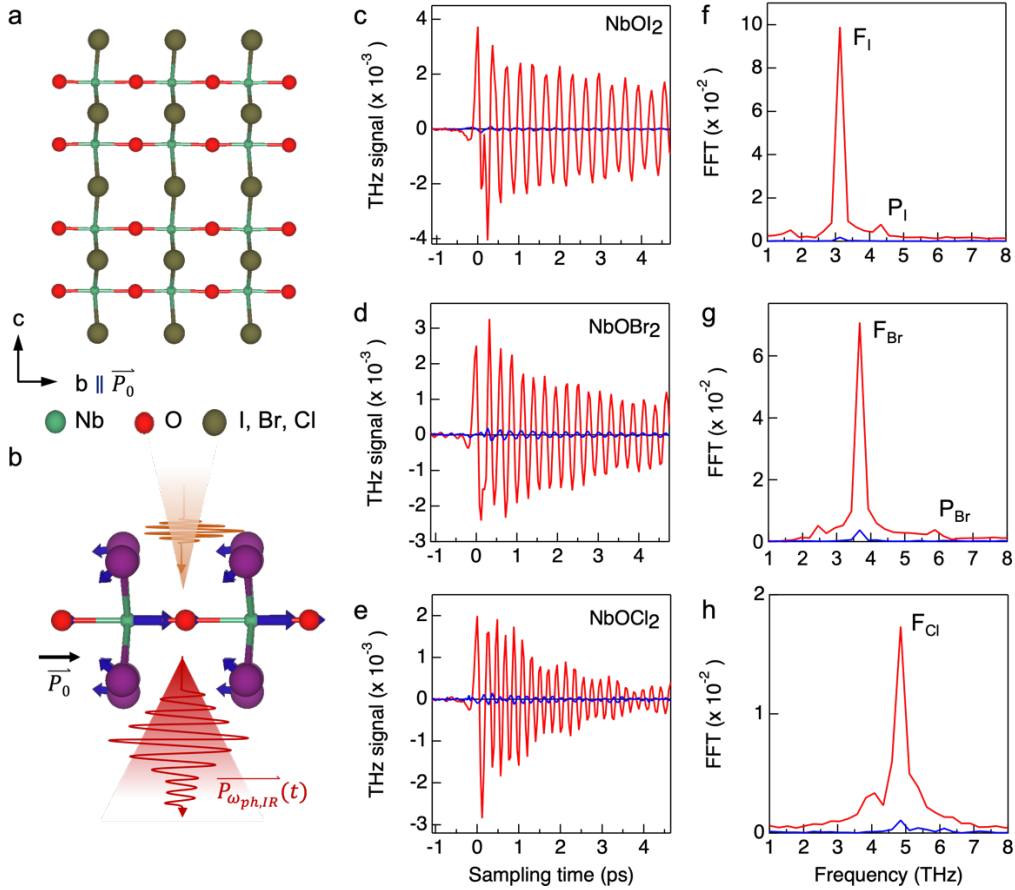

**Fig. 1. Narrowband THz emission and its halogen tuning in vdW ferroelectric $NbOX_2$. a**, In-plane structure of the $NbOX_2$ family (X = I, Br, Cl). They possess robust spontaneous polarization along the polar axis (*b*-axis). **b**, Schematic of THz emission experiment. A NIR femtosecond pulse drives IR-active ferroelectric phonons with an impulsive Raman mechanism, leading to modulation of spontaneous polarization $P_0$ and coherent THz radiation. (**c, d, e**), Time-domain waveform of THz radiation from a (**c**) 330 nm $NbOI_2$, (**d**) 340 nm $NbOBr_2$, and (**e**) 282 nm $NbOCl_2$ flake under 800 nm pulse excitation, collected by a 200 μm GaP EO detector. Signal represented by the red curves are THz radiation parallel to the polar axis, while those by the blue curves are radiation parallel to the non-polar axis. (**f, g, h**), Frequency-domain spectra of a (**f**) $NbOI_2$, (**g**) $NbOBr_2$, and (**h**) $NbOCl_2$ flakes, obtained by Fourier transforming the time-domain traces after 0.7 ps, excluding the optical rectification. $F_X$ denotes emission signal associated with the ferroelectric phonon modes, while $P_X$ denotes a additional emission peak.

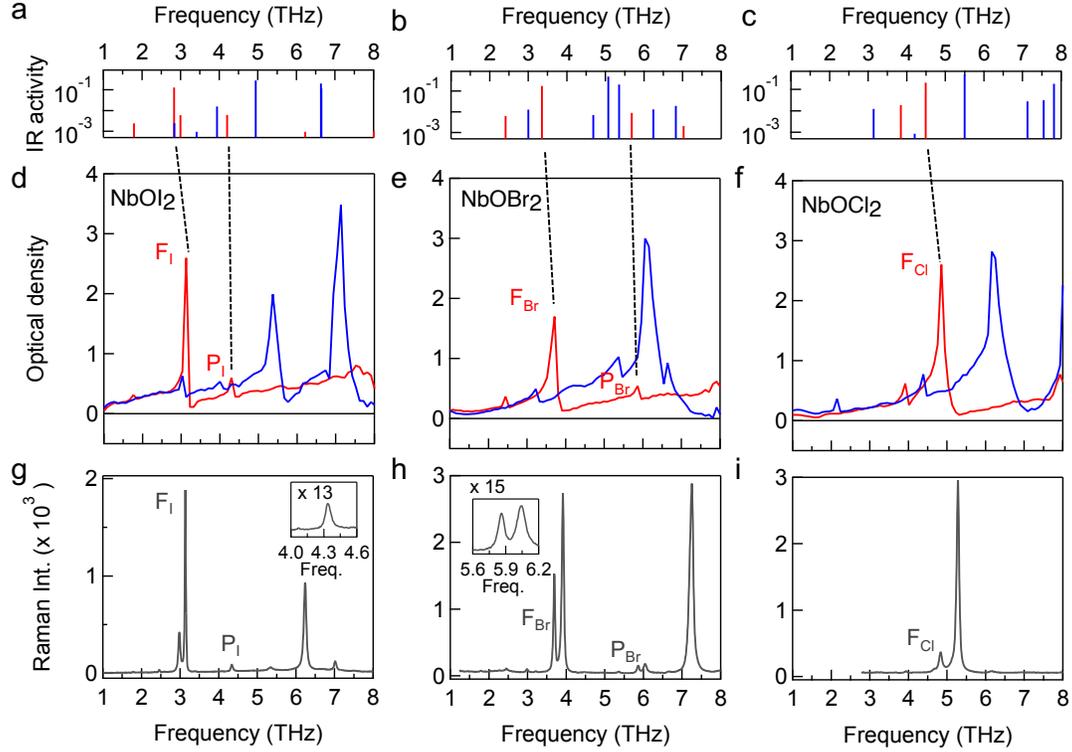

**Fig. 2**. **Holistic IR and Raman spectra of NbOX$_2$.** (**a, b, c**), Calculated IR activities of optical phonons of (a) NbOI$_2$, (b) NbOBr$_2$, and (c) NbOCl$_2$. Red lines represent IR activities of $A$-symmetry phonons, while blue lines represent those of $B$-symmetry phonons. (**d, e, f**), Polarization-resolved THz absorption spectra of a (**d**) 3.70 ± 0.15 μm NbOI$_2$, (**e**) 2.36 ± 0.05 μm NbOBr$_2$, and (**f**) 2.39 ± 0.05 μm NbOCl$_2$ measured by THz-TDS. Red curves show THz absorption parallel to the polar axis, while blue curves represent that parallel to the nonpolar axis. See Supplementary Fig. 3 for the corresponding sample images and time-domain waveforms. (**g, h, i**), Raman spectra of exfoliated (**g**) NbOI$_2$, (**h**) NbOBr$_2$, and (**i**) NbOCl$_2$ flakes excited by a 633 nm laser with incidence angle normal to the vdW plane. Insets of Fig. 2g,h show the relatively weak P$_I$ and P$_{Br}$, respectively. Detailed polarization-angle resolved Raman spectra of each sample are shown in Supplementary Fig. 5, and the data shown in panels g-i are obtained by integrating the polarization-angle resolved spectra.

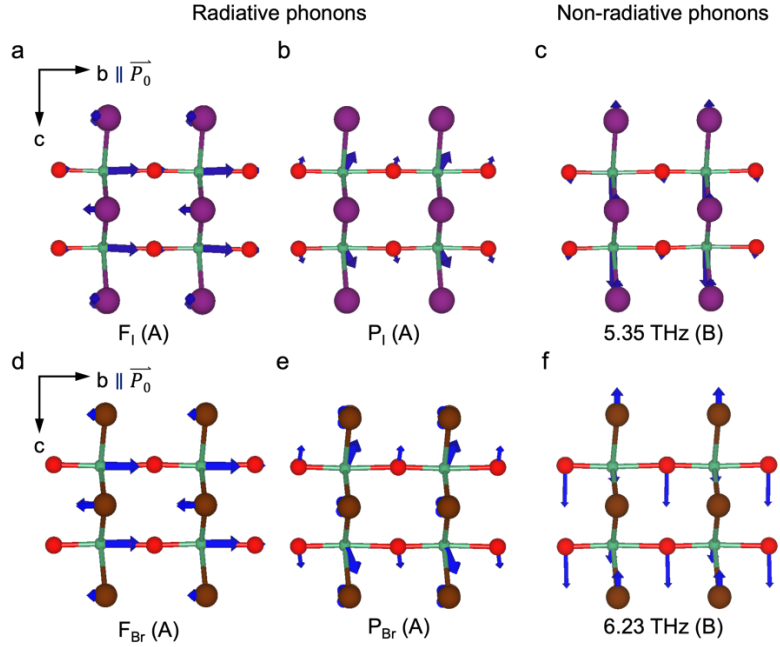

**Fig. 3. Eigenvector analysis.** Eigenvectors of (**a, b, d, e**) *A*-symmetry and (**c, f**) *B*-symmetry phonons viewed perpendicular to the vdW plane (*bc*-plane) of (**a, b, c**) NbOI$_2$ and (**d, e, f**) NbOBr$_2$. The symmetry of the phonon modes is labeled in the parentheses along with the mode labeling. The phonons in (a,b,d,e) become narrowband radiative modes (see text). For radiative modes, the dipole moment induced by the IR-active phonons has net projection on spontaneous polarization $\vec{P_0}$, while this value is zero for non-radiative modes. Nb and O atoms are shown in green and red, while I (Br) atoms are shown with purple (brown).

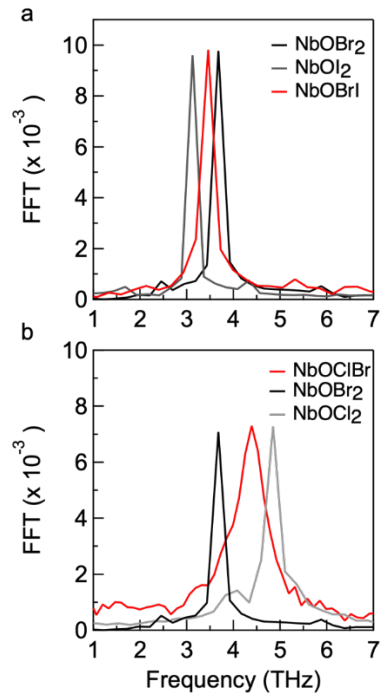

**Fig. 4. Chemical, precise control of THz emission frequency.** Frequency-domain spectra of halide-alloying samples (**a**) NbOBrI and (**b**) NbOClBr. The normalized emission spectra from their parent $NbOX_2$ are reproduced to show continuous tuning of the frequency of the emission peak. See Supplementary Fig. 11 for the corresponding time-domain trace.

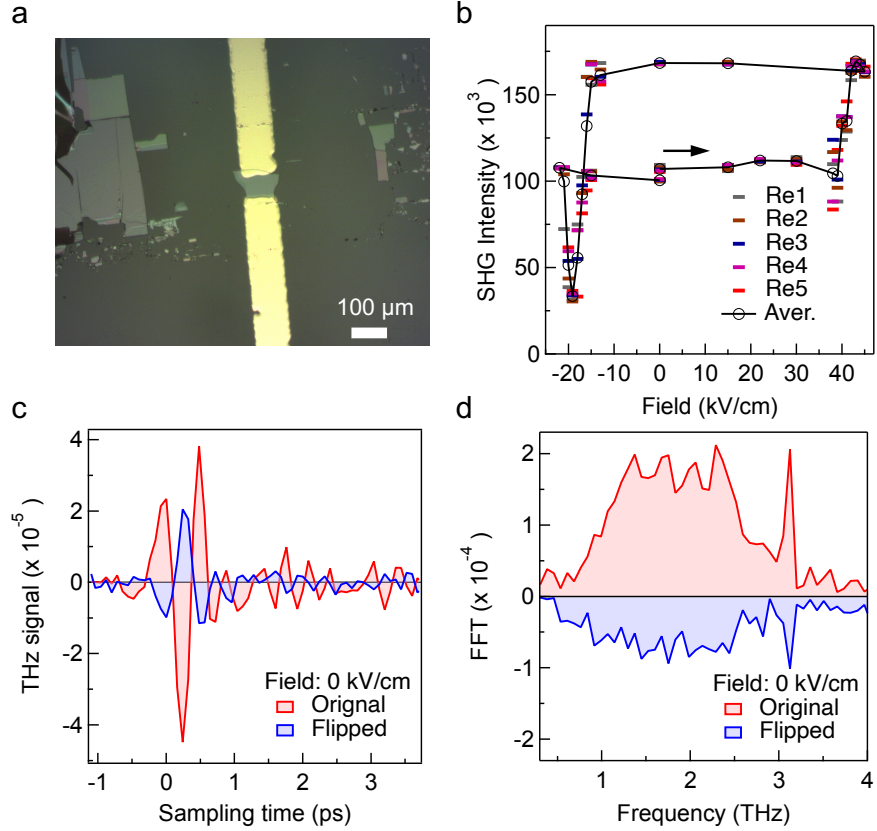

**Fig. 5. Nonvolatile, electrical control of THz emission. a.** Optical image of a NbOI$_2$ flake with two Au/Cr electrodes separated by 70 μm on a fused silica. Thickness of the contacts is designed to be 200 nm to avoid breaks of the electrodes at the edge of the flake, of which the thickness is 120 nm. The electrodes are deposited along *b*-axis of the flake. **b.** Field-dependent SHG measurements, demonstrating switching of the direction of spontaneous polarization and associated change in SHG intensity. Aver. denotes average intensity of five repeats at each applied field. See Methods for details. (**c**, **d**) Electric control of the polarity of coherent THz emission signal in (**c**) time domain and (**d**) frequency domain. The device was trained with -50 kV/cm first to flip the majority of ferroelectric domain and measured without the presence of the externa field.

**Method.**

**Synthesis.**
*Reagents:* The following reagents were used as they were received: niobium powder (Thermo Scientific, Puratronic, 99.99%), niobium oxide (Thermo Scientific, Puratronic, 99.9985%), bromine (Sigma-Aldrich, +99.99%), iodine (Sigma-Aldrich, +99.99%), niobium(V) chloride (Thermo Scientific, Puratronic, 99.999%).

*Synthesis of $NbBr_5$: Caution: Any manipulation involving liquid bromine must be performed inside a well-ventilated fume hood to prevent inhalation of toxic bromine gas.* $NbBr_5$ was synthesized by reaction of the elements and subsequent sublimation described as follows: niobium powder and liquid $Br_2$ in the ratio of 1.1Nb:5Br with a total mass of 6 g were loaded into a fused silica tube (o.d. 12.7 mm and i.d. 10.5 mm). This tube was prepared with a ~ 30 degree bend with the bottom of the tube to the bend 15 cm in length. During the flame sealing of the tube, the bottom was submerged in liquid nitrogen to prevent volatilization of bromine, and the tube was sealed 12 cm from the bend. The bottom of the tube containing the niobium powder was placed inside a tube furnace with the top of the tube outside the furnace angled downward to collect the liquid bromine outside the furnace. The top of the tube was placed in a water bath to ensure the liquid bromine was at room temperature. A blast shield was placed in front of the reaction. The furnace was heated to 275°C, 325°C, and then 500°C dwelling for 18 to 24 hours at each temperature. The product $NbBr_5$ deposits towards the center of the tube, separated from the remaining niobium powder. The tube was opened in a nitrogen-filled glovebox where the product was collected and stored.

*Synthesis of $NbOI_2$:* Single crystals of $NbOI_2$ were prepared using a chemical vapor transport reaction described in detail in reference.[51]

*Synthesis of $NbOCl_2$ and $NbOBr_2$:* Single crystals of $NbOCl_2$ and $NbOBr_2$ were prepared using a chemical vapor transport reaction using the following procedure: Niobium ($NbOCl_2$: 0.0723 g; $NbOBr_2$: 0.0484 g) and niobium(V) oxide ($NbOCl_2$: 0.2070 g; $NbOBr_2$: 0.1385 g) were loaded into a 12.7 mm o.d. and 10.5 mm i.d. fused silica tube. The tube was evacuated and backfilled with argon. Niobium(V) halide ($NbOCl_2$: 0.4357 g of $NbCl_5$; $NbOBr_2$: 0.5281 g of $NbBr_5$) were loaded into the tubes with these reagents exposed to ambient air for only a few seconds to prevent degradation. The tubes were evacuated to a pressure of ~ 30 mtorr and the bottom of the tubes were submerged in liquid nitrogen. The tubes were flame sealed to a length of 12 cm and then placed into a two-zone, tube furnace. The furnace was programmed to the following heating profile: Source side: Heat to 220 °C in 3 h, dwell for 24 h, heat to 500 °C in 24 h, dwell for 102 h, cool to ambient temperature in 6 h. Sink side: Heat to 220 °C in 3 h, dwell for 24 h, heat to 550 °C in 24 h, dwell for 24 h, cool to 450 °C in 6 h, dwell for 72 h, cool to ambient temperature in 6 h. The tubes were opened and stored in air.

*Synthesis of NbOClBr and NbOBrI:* Single crystals of NbOClBr and NbOBrI were prepared identically to the synthesis of $NbOCl_2$ and $NbOBr_2$ except for the following differences: For NbOClBr, 0.0580 g of niobium, 0.1659 g of niobium(V) oxide, 0.1739 g of niobium(V) chloride, and 0.3172 g of niobium(V) bromide were used. For NbOBrI, 0.0835 g of niobium, 0.1179 g of niobium(V) oxide, 0.2244 g of niobium(V) bromide, and 0.2893 g of iodine were used with the iodine added at the same time as the $NbBr_5$. The furnace was programmed to the following heating profile: Source side: Heat to 220 °C in 3 h, dwell for 24 h, heat to 500 °C in 24 h, dwell for 126 h, cool to ambient temperature in 6 h. Sink side: Heat to 220 °C in 3 h, dwell for 24 h, heat to 550 °C in 24 h, dwell for 48 h, cool to 450 °C in 6 h, dwell for 72 h, cool to ambient temperature in 6 h.

**Preparation of exfoliated flakes**
The crystals were mounted on Scotch tape and cleaved several times to expose fresh and flat surface. The tape with flakes were then placed on top of a fused silica substrate with thickness of 1 mm (Corning 7980) and heated at 70 °C for 10 mins on a hotplate. The substrate with flakes was then cooled down to room temperature prior removal of the tape. The flakes were then successfully transferred to the substrate. The thickness of each flake was determined by atomic force microscopy (AFM).

**Terahertz experiment**
THz measurements were performed using our home-built far-field THz setup.[52] Briefly, the output of Ti:sapphire regenerative amplifier (RA) with a pulse duration of 30 fs, a repetition rate of 10 kHz, and a wavelength centered at 800 nm (Coherent, Legend) was separated into two beams for pump and sampling. For THz emission, the pump beam was chopped at 5 kHz, and then passed an achromatic half-wave plate and was focused onto a $NbOX_2$ sample from the substrate side with the pump spot size being around 50 μm (1/e radius). The excitation fluence was 7.32 mJ cm$^{-2}$ for most of $NbOX_2$ samples except for $NbOBr_2$, which was pumped with a fluence of 10.5 mJ cm$^{-2}$. No observable damage due to excitation under eye check with a microscope. The polarization of the pump beam was controlled to be parallel to the detection axis of the electro-optic (EO) detector. The orientation of the samples was manually adjusted to align with the polarization of the pump beam. The THz emission signal was collected using a parabolic mirror and passed through a high-resistivity Si wafer and a high-density polyethylene plate. Two THz polarizers (PureWave) were placed thereafter to set the detection polarization parallel to the EO detector. The THz signal was then measured using a 200-μm <110> GaP crystal with the sampling beam via EO sampling, except for Fig. 5, which was recorded using a <110> 1-mm ZnTe. The setup was purged with $N_2$ gas during the measurements, and the samples were placed in $N_2$ at room-temperature. The EO signal encoded in the sampling beam then went through a quarter-wave plate and subsequently a Wollaston prism to have two linear polarized components be separated. The signal was then recorded using a balanced detector, box-car integrator, and a data acquisition card.

The transmittance ratios of a pair of wire grid polarizers (PW005-012, PureWave) for orthogonal polarized THz wave in power spectrum at 3.1, 3.7, and 4.9 THz were estimated to be around $10^{-6}$, $10^{-5}$, and 2 x $10^{-5}$, based open access data from PureWave. The leakage narrow-band power signal from the polar axis of $NbOX_2$ was estimated to be 2 x $10^{-4}$, 2 x $10^{-3}$, 2 x $10^{-3}$, for $NbOI_2$, $NbOBr_2$, and $NbOCl_2$, respectively. The higher transmittance ratio is attributed to imperfect radiation incidence deviated from normal incidence.

For THz-TDS measurements, the broadband THz probe light was generated with a two-color air plasma method by the pump beam through an alternative beam path. This THz probe was collimated and focused onto the samples mounted on precision pinholes (Thorlabs) with diameter of 1 mm ($NbOI_2$, $NbOBr_2$) or 500 μm ($NbOCl_2$) that define sampling area. Reference THz signal was taken with an empty pinhole. The absorption spectra (represent in optical density) were obtained by calculating $OD = -\log \frac{E^2_{\text{Sample}}}{E^2_{\text{Reference}}}$ after Fourier transformation. Prior to THz-TDS and THz emission measurements, the crystal orientation was first determined using *in situ* white-light polarimetry equipped inside the THz setup.

We calculate the thickness of each sample by analyzing interference patterns of the *c*-axis reflectance spectra of the samples with reported or measured refractive index (Supplementary Fig 14).[53]

**Polarization-angle resolved Raman measurements**

The low-frequency Raman scattering measurements were carried on a home-built setup with a Nikon TE-300 inverted microscopy. A 633 nm HeNe laser first passed a laser line filter, and a linear polarizer and was filtered by a 90/10 beamsplitter filter (Ondax), then sent through a half-wave plate to control the angle of the incident polarization, and into the microscope objective (40X, NA = 0.6). The incident light was focused onto $NbOX_2$ flakes placed on $Si/SiO_2$ substrates secured in a vacuum environment (Oxford MicroStat Hi-Res2). The Raman scattering light was collected through the same objective and the same half-wave plate, which projects the parallel-polarized Raman scattering onto the polarization axis of the incident laser and the cross-polarized Raman signal onto the orthogonal axis. The Raman signal was subsequently sent through two-notch filters (Ondax) to filter out the Rayleigh scattering laser line. Then, the Raman signal was focused onto the entrance slit of a spectrometer (Princeton Instruments HRS-300) with a 2400 gr/mm holographic grating that dispersed the spectrum onto a LN2 cooled (charge-coupled device) CCD camera (Princeton Instruments LN400/B). The Raman shift was calibrated by a dual Hg/He atomic emission lamp.

**Visible-light reflectance polarimetry**

An external visible imaging and polarimetry system were coupled to the THz-TDS set-up, which was used to remotely determine the crystal orientation of each flake prior to THz experiments and estimate the thickness from etalon interference. White light (fiber-coupled illuminator, Thorlabs) was sent through a through hole on a parabolic mirror. The achromatic half-wave plate described above was used to control the polarization of the incident white light. The 1.5-cm-focal-length lens was used to focus the white light onto the sample and collect the reflected light. When imaging, an additional lens was inserted to make a Köhler-type illumination. The collected reflected light was sent to an imaging camera or a fiber end with a resettable mirror. The other end of the fiber was coupled to a $LN_2$-cooled CCD camera equipped with a monochromator (Princeton), which measures the polarization-dependent reflectivity.

**Field dependent measurements**

For device fabrication, a $NbOI_2$ flake was first transferred to a fused silica substrate following the procedures above. Thickness and crystal orientation were first identified with an AFM and polarization-angle resolved second-harmonic generation (pSHG) measurement, respectively. Residual flakes surrounding the target flakes were picked up by a dry-transfer technique with polycarbonate stamp prior to contact fabrication.[54] To avoid exposure to water, the contacts were made of a two-layer metal film of Cr/Au (3 nm/200 nm) and deposited with a customed shadow mask (Stencils Unlimited).

For the field-dependent pSHG measurement, the sample was mounted in a cryostat and maintained in vacuum throughout the measurement, and wired to an external source meter (Keithley 2400).

We carried out pSHG with 80 MHz femtosecond pulses (800 nm, 100 fs) generated by a Ti: sapphire oscillator (Mira 900, Coherent). The pump beam was first sent through a linear polarizer, subsequently reflected by a 650-nm short-pass dichroic mirror, and passed an achromatic half-wave plate. The pump beam was then focused onto the sample with a 50x objective. The SHG signal was collected with the same objective, passed the same achromatic half-wave plate, and transmitted through the short-pass dichroic mirror. The SHG signal then passed a 700-nm short-pass filter and was collected by a photomultiplier tube. For the field-dependent THz emission measurement, the same sample was kept in the vacuum and placed into the THz emission setup.

At each applied field, we repeated 5 scans of SHG polarimetry to reveal the dynamics of ferroelectric phase transition. Each data spot shown in Fig. 5b represents an integrated intensity of each scan of SHG polarimetry, and Aver. represents an average of the integrated intensity of the 5 scans. The current at each applied field was *in situ* collected and represented in Supplementary Fig. 12

**First-principles calculation**
Phonon and IR/Raman spectra were computed from first-principles density functional theory (DFT). Electronic structures and structural relaxations used the Perdew-Burke-Ernzerhof (PBE) functional[55] as implemented in VASP.[56,57] Calculations employed the primitive cell of $NbOX_2$ (X = Cl, Br, I) with a plane-wave cutoff of 500 eV and a Γ-centered 6×6×6 k-point mesh. Atomic positions were relaxed until residual forces were below $1\times10^{-3}$ eV Å$^{-1}$.

Phonons were obtained via the finite-displacement (supercell "frozen phonon") method using Phonopy[58,59] with 3×3×3 supercells. Non-analytical term corrections (LO-TO splitting) were included using Born effective charge tensors and the electronic dielectric tensor, both evaluated by density functional perturbation theory (DFPT).[60] IR activities were derived by finite differences of the macroscopic dielectric tensor with respect to mode-projected displacements.[61,62]

**Phonon symmetry analysis**
The symmetry of phonons of the $NbOX_2$ family can be categorized to either A or B as an irreducible representation under the *C*2 space group. The corresponding character table was documented in Supplementary Table 3. From this table, the phonons belong to A symmetry must have IR activities parallel to the polar axis, which is parallel to the C2 principal rotation axis, while those belong to B symmetry must have IR activities along both non-polar and out-of-plane directions. For the current study, we limit our discussion to the in-plane IR activities. The symmetry of $F_I$, $F_{Br}$, $F_{Cl}$, $P_I$, and $P_{Br}$ were determined to be A by Raman tensor analysis shown in Supplementary Fig. 6, agreeing with the calculated phonon symmetry and the experimentally observed IR absorption spectra shown in Fig. 2d-f.

**Data Availability Statement**
The data within this paper are available upon reasonable request.

**ACKNOWLEDGEMENTS.** Experiments on THz emission were supported by the US Army Research Office under grant number W911NF-23-1-0056. Material synthesis and first-principles calculations were supported by the Materials Science and Engineering Research Center (MRSEC) on Precision Assembly of Quantum Materials (PAQM) through NSF award DMR-2011738. Structural, Optical, Raman, and SHG characterizations were supported by the Air Force Office of Scientific Research under award number FA9550-22-1-0389. This research utilized instrumentation provided by the Programmable Quantum Materials, an Energy Frontier Research Center funded by the U.S. Department of Energy (DOE), Office of Science, Basic Energy Sciences (BES), under award DE-SC0019443. CYH acknowledges support from the Taiwan-Columbia scholarship funded by the Ministry of Education of Taiwan and Columbia University. MLF was supported by the National Science Foundation Graduate Research Fellowship Program (DGE-2036197). MF and XYZ acknowledge support for collaboration on theoretical analysis by the Max Planck – New York City Center for Non-Equilibrium Quantum Phenomena.

**Author contributions**
C.-Y.H., T.H., X.R., and X.Z. conceived this project; D.G.C synthesized the crystals under the supervision of X.R.; C.-Y.H prepared the samples; C.-Y.H and T.H carried out the optical measurements; Z.-H.C. performed first-principles calculation under the supervision of D.R.R.; C.-Y.H characterized the crystals structure and composition with assistance by M.L.F. and D.G.C.; C.-Y.H. fabricated the device with assistance by Y.L.; D.X. measured the dielectric function of $NbOBr_2$ under the supervision of M.E.D.; M.F. proposed theory framework for narrow-band THz emission. The manuscript was prepared by C.-Y.H., T.H., X.R., and X.Z in consultation with all other authors. All the authors read and comment on the paper.

**Competing Interests**
The authors declare no competing interests.

**Supplementary Information for**

**Tunable Narrowband Coherent Teraertz Emission from van der Waals Ferroelectrics**

Chun-Ying Huang[1,*], Taketo Handa[1,*], Daniel G. Chica[1,*], Zhihao Cui[1], Ding Xu[1], Jeongheon Choe[1], Yiliu Li[1], Margalit L. Feuer[1], Milan E. Delor[1], Michael Fechner[2], David R. Reichman[1], Xavier Roy[1,†], & Xiaoyang Zhu[1,†]

[1]Department of Chemistry, Columbia University, New York, NY 10027, USA
[2]Max Planck Institute for the Structure and Dynamics of Matter, Hamburg, Germany
[*]These authors contribute equally
[†]To whom correspondence should be addressed: Xavier Roy: xr2114@columbia.edu; Xiaoyang Zhu: xyzhu@columbia.edu.

**Supplementary Method.**

**Determination of the dielectric function of NbOBr$_2$.**
The dielectric function of NbOBr$_2$ was extracted from angle-resolved reflectance spectroscopy. The sample thickness was first determined by atomic force microscopy. Broadband illumination from a halogen lamp was directed onto the sample, and the reflected signal was collected through a reflective microscope equipped with an oil-immersion objective (NA=1.45). An adjustable slit and a linear polarizer were positioned at the back focal plane to select the angle-resolved reflectance along the principal crystal axes. The slit-filtered light was dispersed by a home-built prism spectrometer, where the horizontal axis corresponds to photon energy and the vertical axis represents reflection angle. The resulting angle-resolved reflectance spectra were analyzed using transfer-matrix-method fitting. The dielectric function was parameterized by a multioscillator Tauc-Lorentz model, globally fitted to several angle-resolved spectra collected from flakes of different thicknesses, enabling consistent extraction of optical constants across the measured spectral range.

**Powder X-ray diffraction (PXRD) measurements.**
Diffraction patterns were collected under an N$_2$ environment using a PANanalytical Aeris diffractometer with a Cu Kα X-ray source energized to 40 kV and 15 mA. Samples were placed on a Si-zero background holder which was spun during the collection time.

**Energy-Dispersive X-Ray Spectroscopy (EDX).**
EDX measurements were performed with JEOL JCM-7000 NeoScope scanning electron microscopy (SEM) operated at an accelerating voltage 15 kV. Bulk crystals of NbOBrCl and NbOBrI were attached to a sample holder with conductive carbon tape and cleaved with Scotch tape right before placed under vacuum.

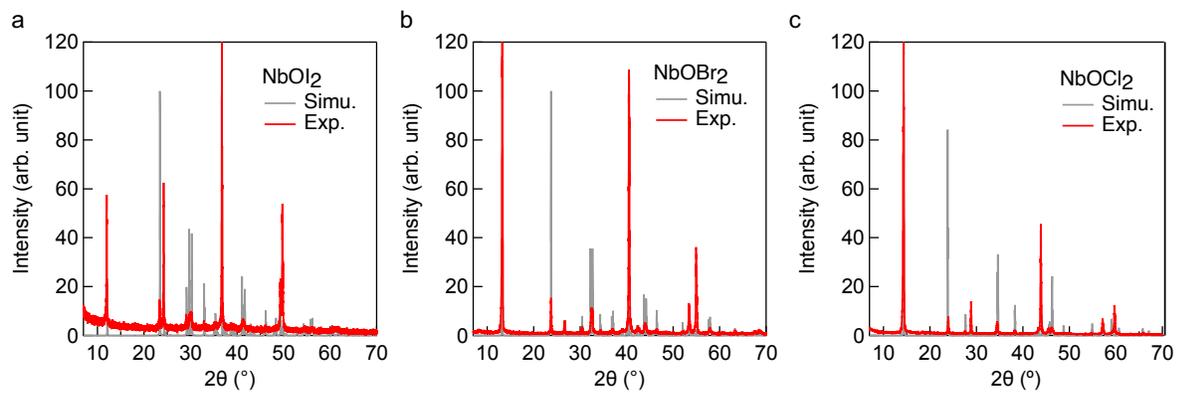

**Supplementary Fig. 1.** Powder X-ray diffraction spectra of (a) $NbOI_2$, (b) $NbOBr_2$, and (c) $NbOCl_2$. Simu. represents simulated patterns of the structures from open sources.

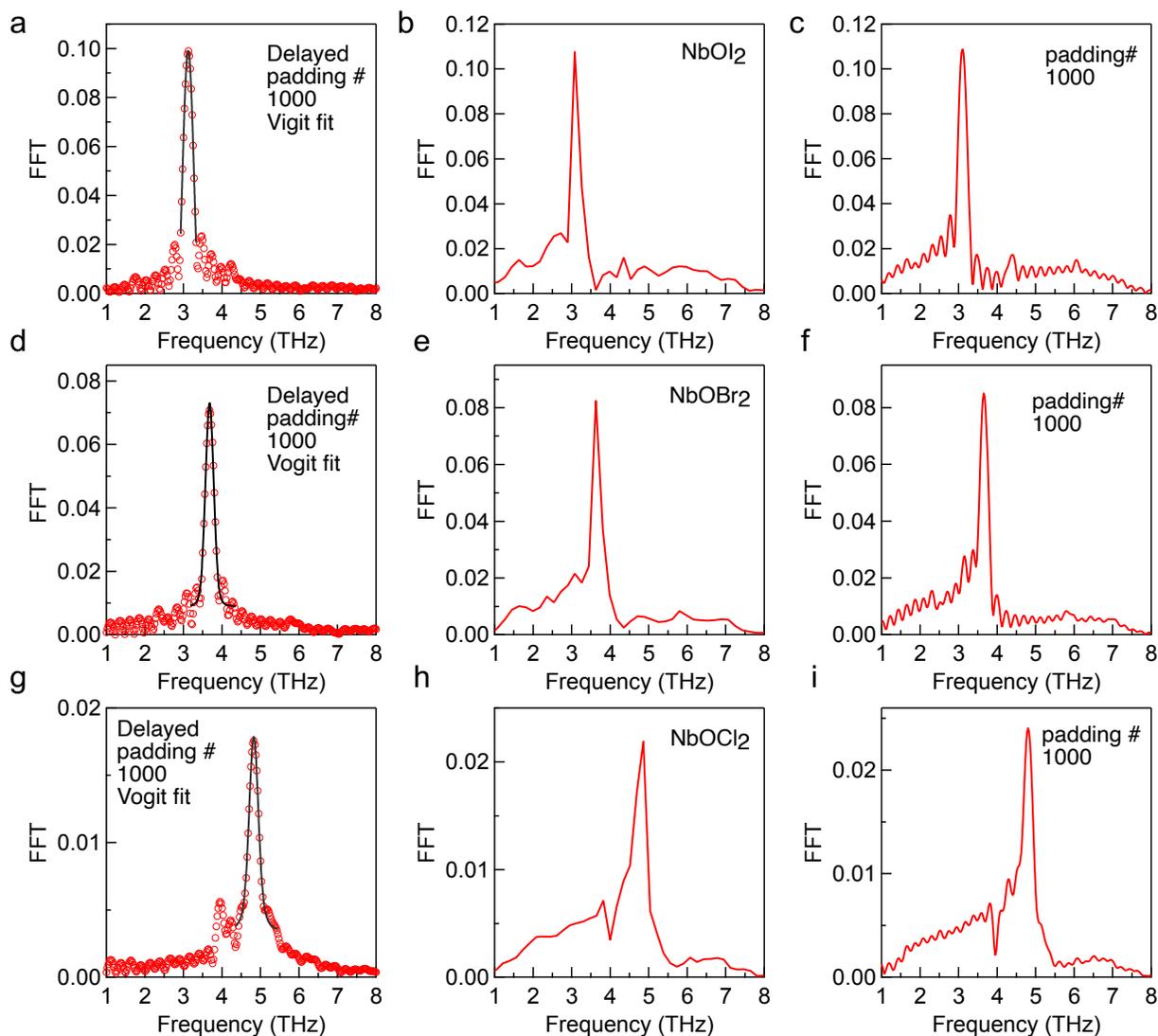

**Supplementary Fig. 2.** Extend frequency-domain spectra of (**a-c**) NbOI$_2$, (**d-f**) NbOBr$_2$, and (**g-i**) NbOCl$_2$. Panel **a**, **d**, **g** show a Vogit fit to each identified ferron peak (F$_I$, F$_{Br}$, F$_{Cl}$) after a sufficient padding number that provides better frequency resolution to define center frequency. Panel **b**, **e**, **h** show FT of complete time-domain traces without adding padding number to recover the phase information. Panel **c**, **f**, **i** show FT spectrum with additional padding number to clearly show the dip feature in each emission spectrum. These dip features correspond to reabsorption of IR-active TO phonons at (**c**) 1.8 THz for NbOI$_2$, (**f**) 2.4 THz for NbOBr$_2$, and (**i**) 4 THz for NbOCl$_2$.

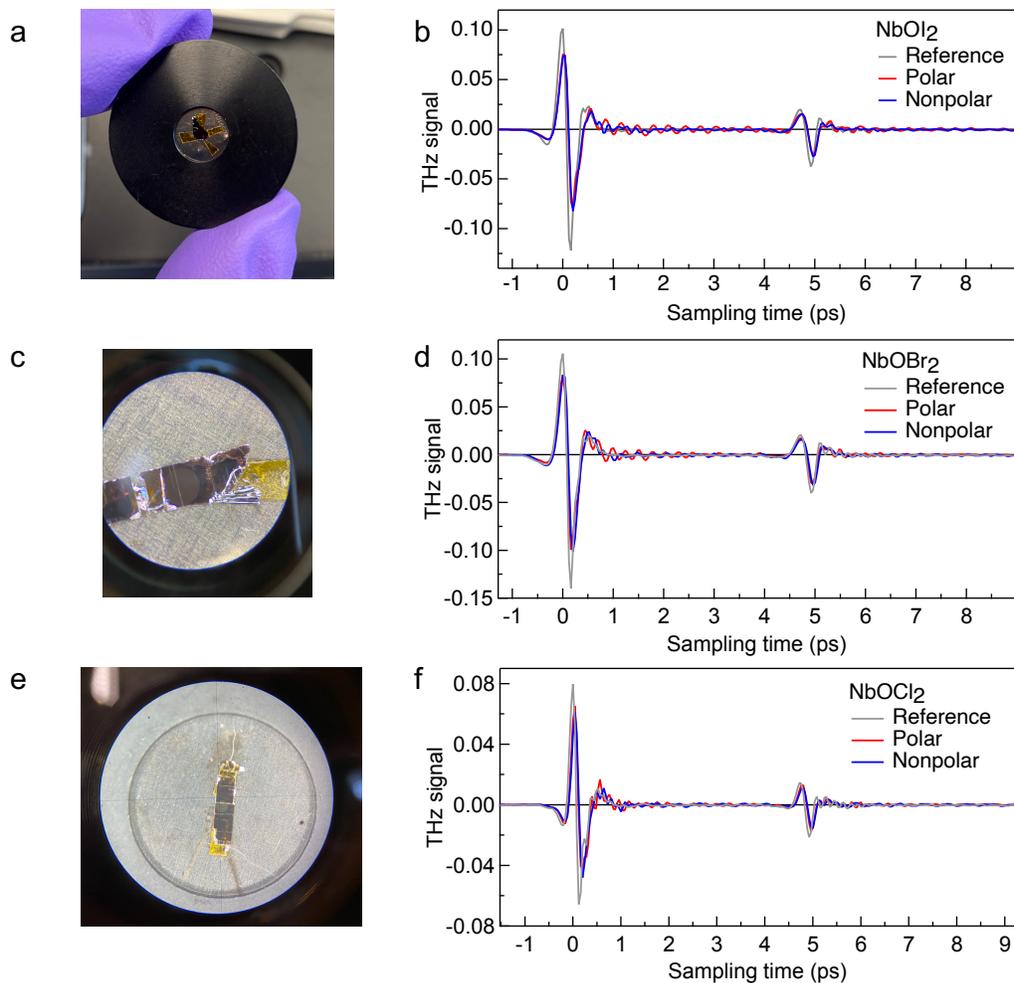

**Supplementary Fig. 3.** Sample pictures and corresponding THz-TDS in time domain covering full sampling time of (**a**, **b**) NbOI$_2$, (**c**, **d**) NbOBr$_2$, and (**e**, **f**) NbOCl$_2$. NbOI$_2$ and NbOBr$_2$ were mounted on a 1-mm pinhole, while NbOCl$_2$ was mounted on a 500-µm pinhole due to limited crystal size. An echoing signal arrives at around 4.5ps for each measurement.

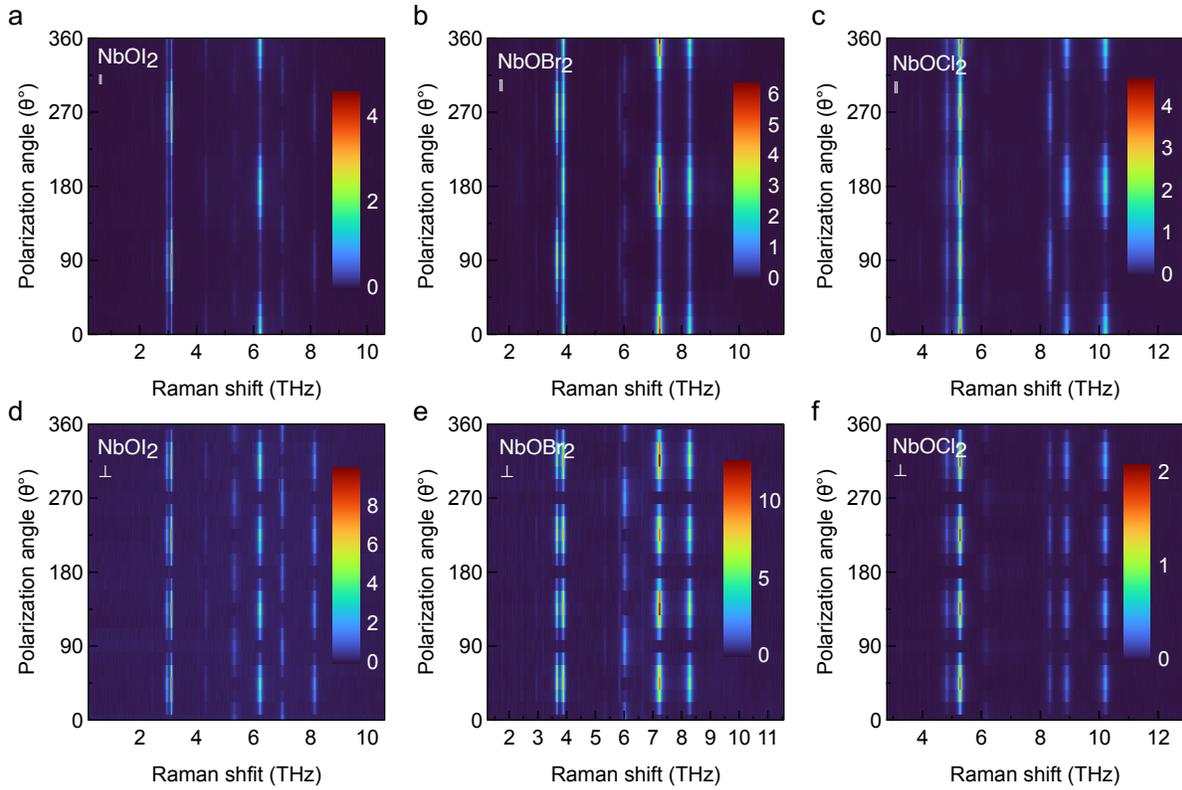

**Supplementary Fig. 4.** Polarization-angle resolved Raman spectra of (**a, d**) NbOI$_2$, (**b, e**) NbOBr$_2$, and (**c, f**) NbOCl$_2$. (**a, b, c**) Parallel polarization denoted with ∥. (**d, e, f**) Cross polarization denoted with ⊥. Color scale represents Raman intensity. 0° polarization angle is aligned with c-axis (non-polar axis) for all Raman measurements.

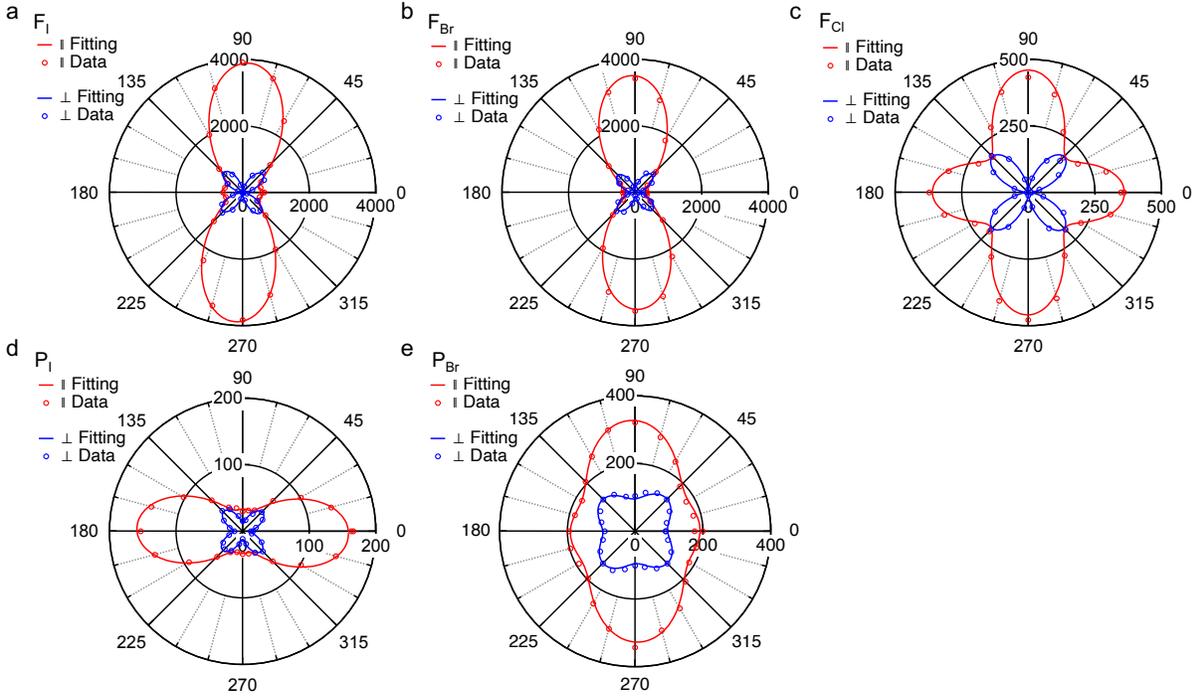

**Supplementary Fig. 5.** Raman tensor analysis for the observed radiative phonons of (**a**, **d**) NbOI$_2$, (**b**, **e**) NbOBr$_2$, and (**c**) NbOCl$_2$. We fit the polarization-angle dependence of both the parallel (∥) and cross (⊥) polarized Raman intensity with *A* symmetry under *C2* space group, which follows:[1–3]

$$I_{A, \parallel} = A(|c|^2 \cos^4 \theta + |b|^2 \sin^4 \theta + 2|b||c| \sin^2 \theta \cos^2 \theta \cos \varphi_{c\text{-}b}) + bg \quad (1)$$
$$I_{A, \perp} = A/4(|c|^2 + |b|^2 - 2|b||c| \cos \varphi_{c\text{-}b}) \sin^2 2\theta + bg \quad (2)$$

For phonons with B symmetry, the Raman intensity *vs* polarization angle is described by the equations below:

$$I_{B, \parallel} = A|f|^2 \sin^2 2\theta + bg \quad (3)$$
$$I_{B, \perp} = A|f|^2 \cos^2 2\theta + bg \quad (4)$$

An isotropic constant is included to account for non-zero background intensity.

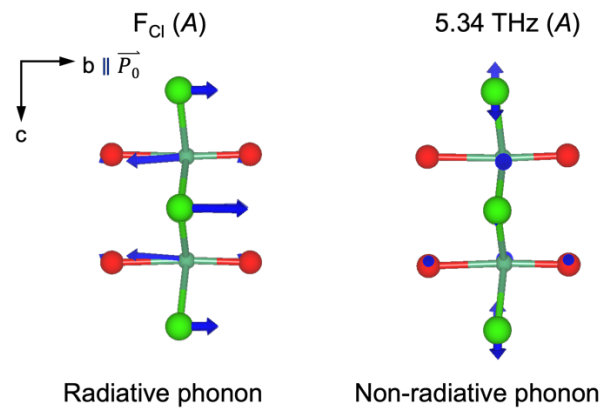

**Supplementary Fig. 6.** Selected calculated eigenvectors of NbOCl$_2$ phonons. Both eigenvectors belong to *A* symmetry. Absence of dipole change can be observed from the 5.34 THz one, resulting in no radiation in the THz emission spectrum.

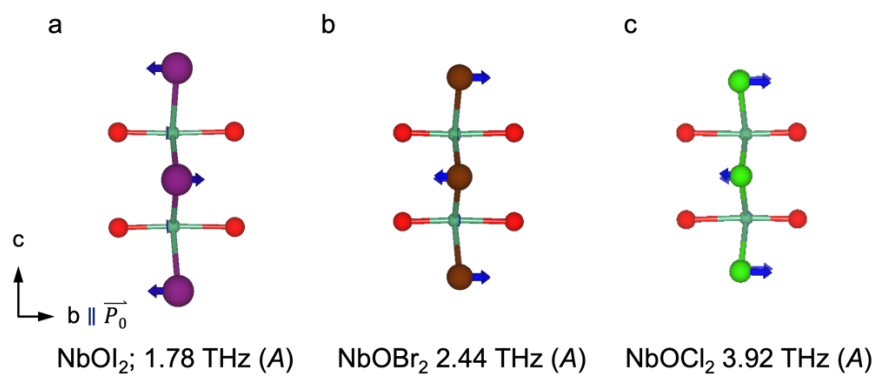

**Supplementary Fig. 7.** Energetically low-lying phonon modes with *A* symmetry of (**a**) NbOI$_2$, (**b**) NbOBr$_2$, and (**c**) NbOCl$_2$ No presence of direct modulation on the dipole moment leading to the spontaneous polarization, resulting in no radiation in the THz emission spectrum.

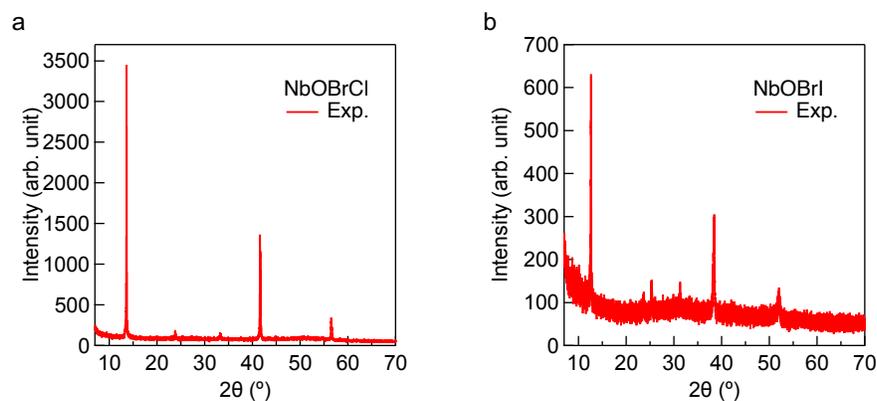

**Supplementary Fig. 8.** Powder X-ray diffraction spectra of (**a**) NbOClBr and (**b**) NbOBrI.

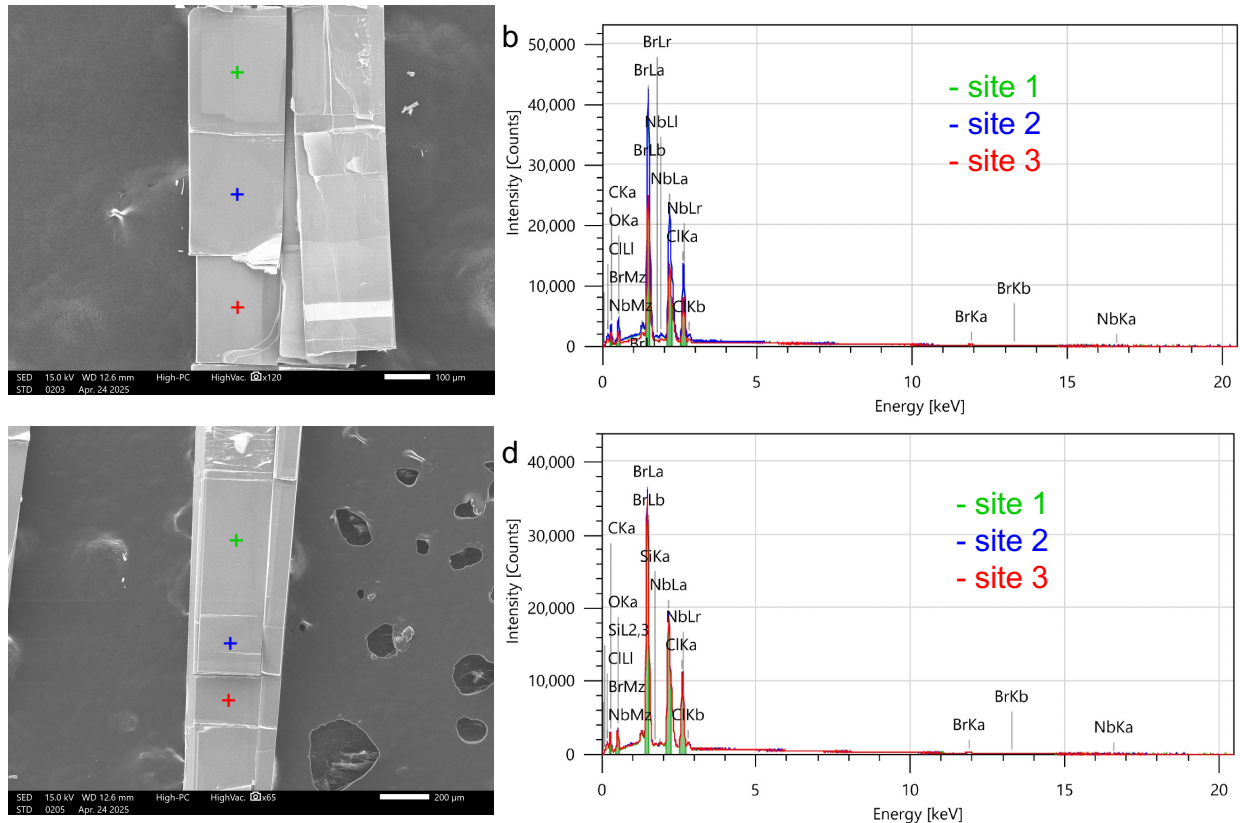

**Supplementary Fig. 9.** Scanning electron microscopy (SEM) images of NbOClBr crystals and corresponding energy dispersive X-ray spectroscopy (EDX) spectra. Color labels were placed on SEM images to show the locations taken for EDX measurements.

**Supplementary Table 1.** Elemental composition of Nb, Cl, and Br, determined from EDX from multiple spots of two NbOClBr crystals.

| Crystal # | Spot # | Nb (Mass %) | Cl (Mass %) | Br (Mass %) |
|---|---|---|---|---|
| 1 | 1 | 45.38 | 15.78 | 38.84 |
|   | 2 | 45.64 | 15.73 | 38.63 |
|   | 3 | 45.52 | 15.92 | 38.56 |
| 2 | 1 | 45.44 | 15.57 | 38.99 |
|   | 2 | 45.49 | 15.55 | 38.97 |
|   | 3 | 45.45 | 15.52 | 39.03 |

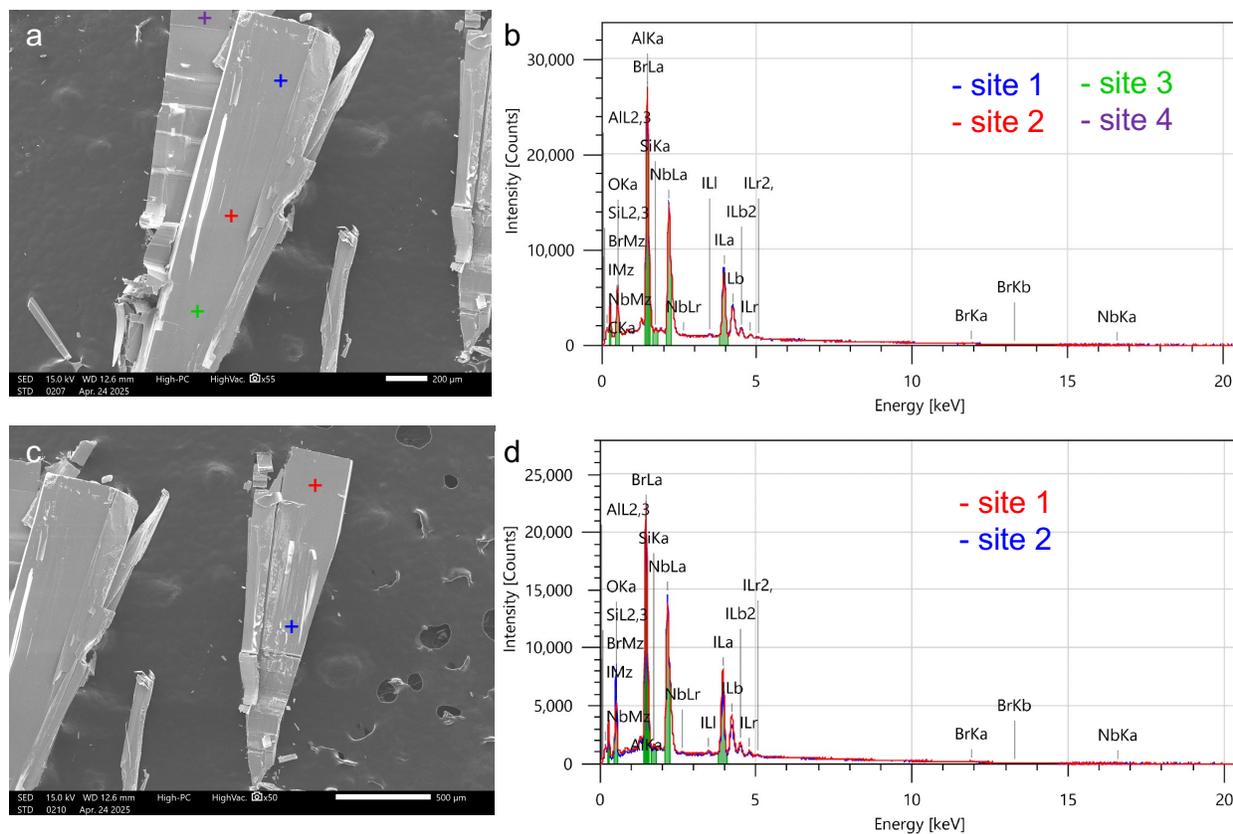

**Supplementary Fig. 10.** Scanning electron microscopy (SEM) images of NbOBrI crystals and corresponding energy dispersive X-ray spectroscopy (EDX) spectra. Color labels were placed on SEM images to show the locations taken for EDX measurements.

**Supplementary Table 2.** Elemental composition of Nb, Br, and I, determined from EDX from multiple spots of two NbOBrI crystals.

| Crystal # | Spot # | Nb (Mass %) | Br (Mass %) | I (Mass %) |
|---|---|---|---|---|
| 1 | 1 | 33.14 | 33.02 | 33.84 |
|   | 2 | 33.07 | 30.43 | 36.49 |
|   | 3 | 32.75 | 30.82 | 36.43 |
|   | 4 | 32.44 | 31.24 | 36.31 |
| 2 | 1 | 32.59 | 29.00 | 38.41 |
|   | 2 | 38.55 | 28.44 | 35.71 |

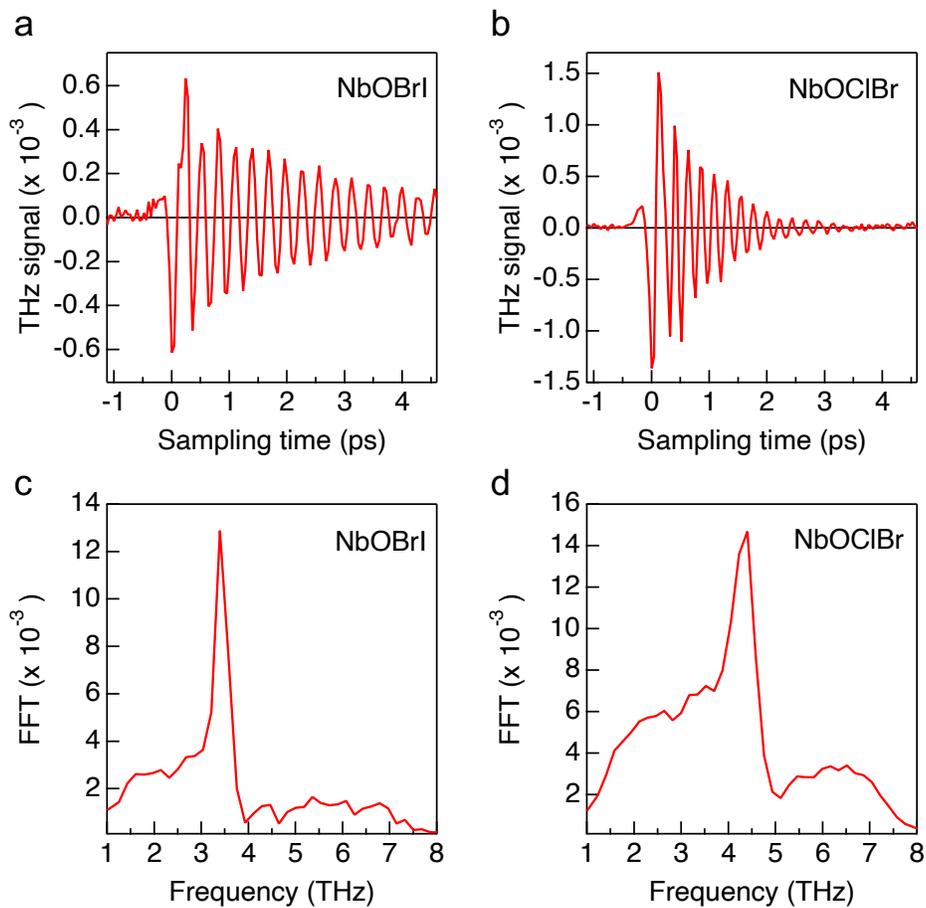

**Supplementary Fig. 11.** THz emission spectra of NbOBrI and NbOClBr in (**a**, **b**) time-domain and (**c**, **d**) frequency domain.

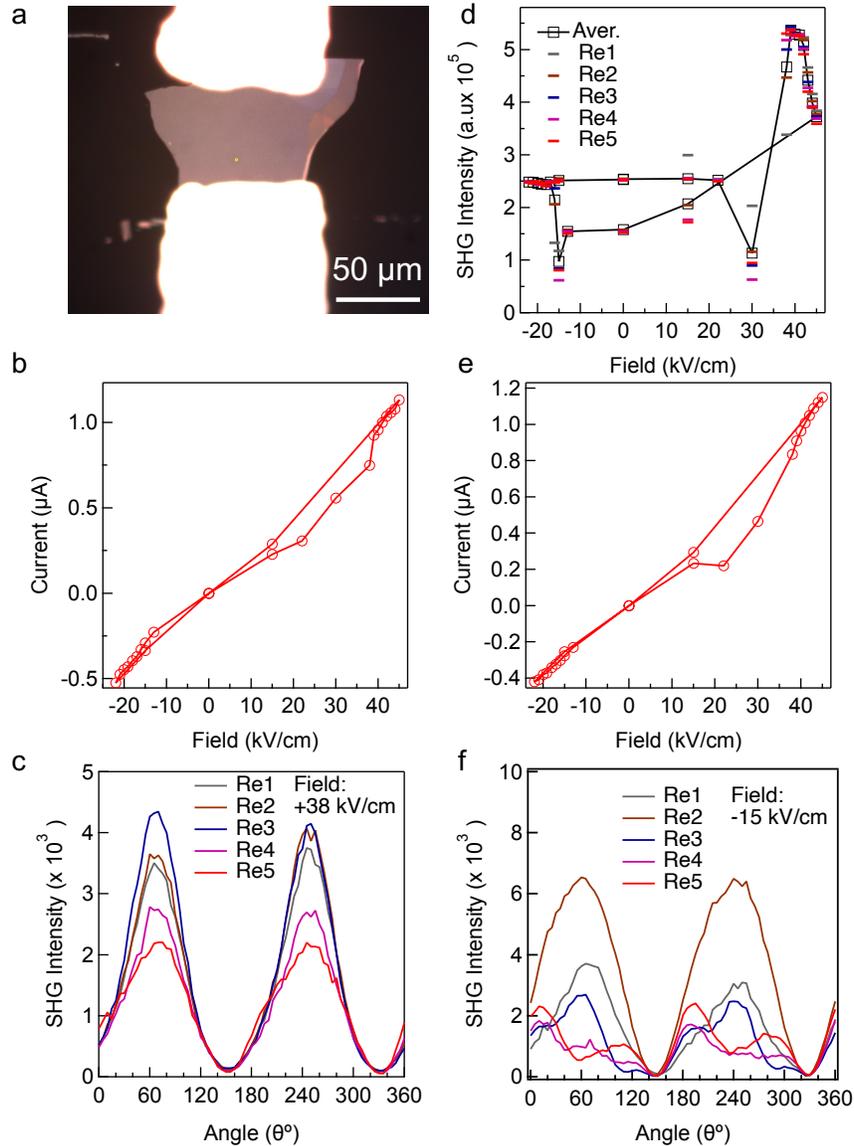

**Supplementary Fig. 12. a**. Additional device picture. **b**. Profile of current vs applied field of the first SHG scan shown in **Fig. 5b**. **c**. SHG polarimetry with an external field of + 38 kV/cm of the first scan, where real-time change in SHG intensity was observed among 5 polarimetry scans. **d**. Second field-dependent SHG scan. **e**. Profile of current vs applied field of the second SHG scan shown in (**b**). **f**. SHG polarimetry with an external field of - 15 kV/cm of the second SHG scan, where real-time change in SHG intensity and polarization-angle dependence were observed among 5 polarimetry scans. The real-time change observed above indicates a slow process of ferroelectric domain switching at the threshold field.

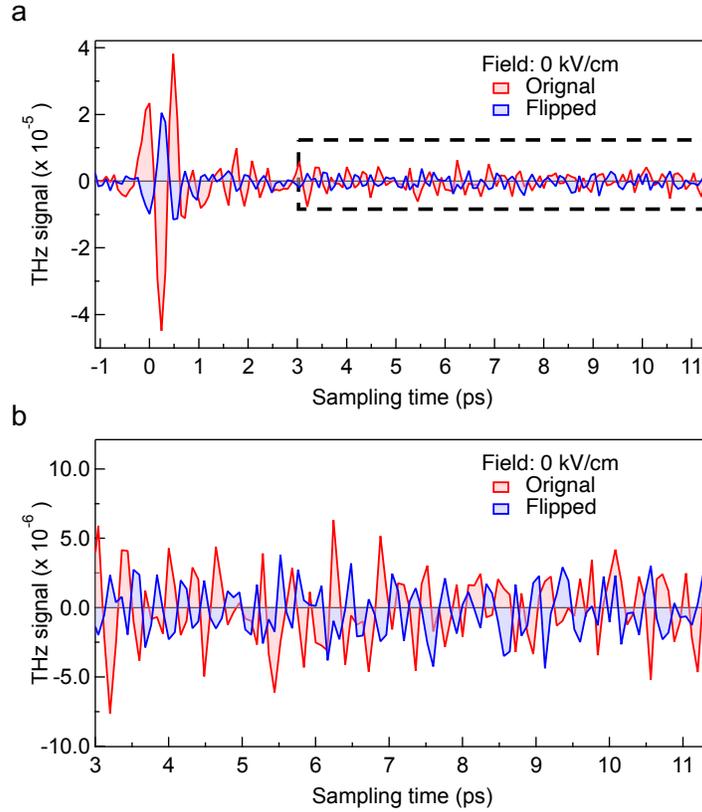

**Supplementary Fig. 13. a**. Time-domain traces of the NbOI$_2$ flake (ref) before and (blue) after high-field training. **b**. Zoom-in of the coherent THz emission after 3 ps, showing the opposite phase of the waveforms between the ferroelectric domain switching.

**Supplementary Table 3.** Character table of *C2* space group[3]

| System | Point group | Irreducible representation | Raman tensors | Linear function |
|---|---|---|---|---|
| Monoclinic | C$_2$ | A | $\begin{pmatrix} a & & d \\ & b & \\ d & & c \end{pmatrix}$ | y |
| | | B | $\begin{pmatrix} & e & \\ e & & f \\ & f & \end{pmatrix}$ | x, z |

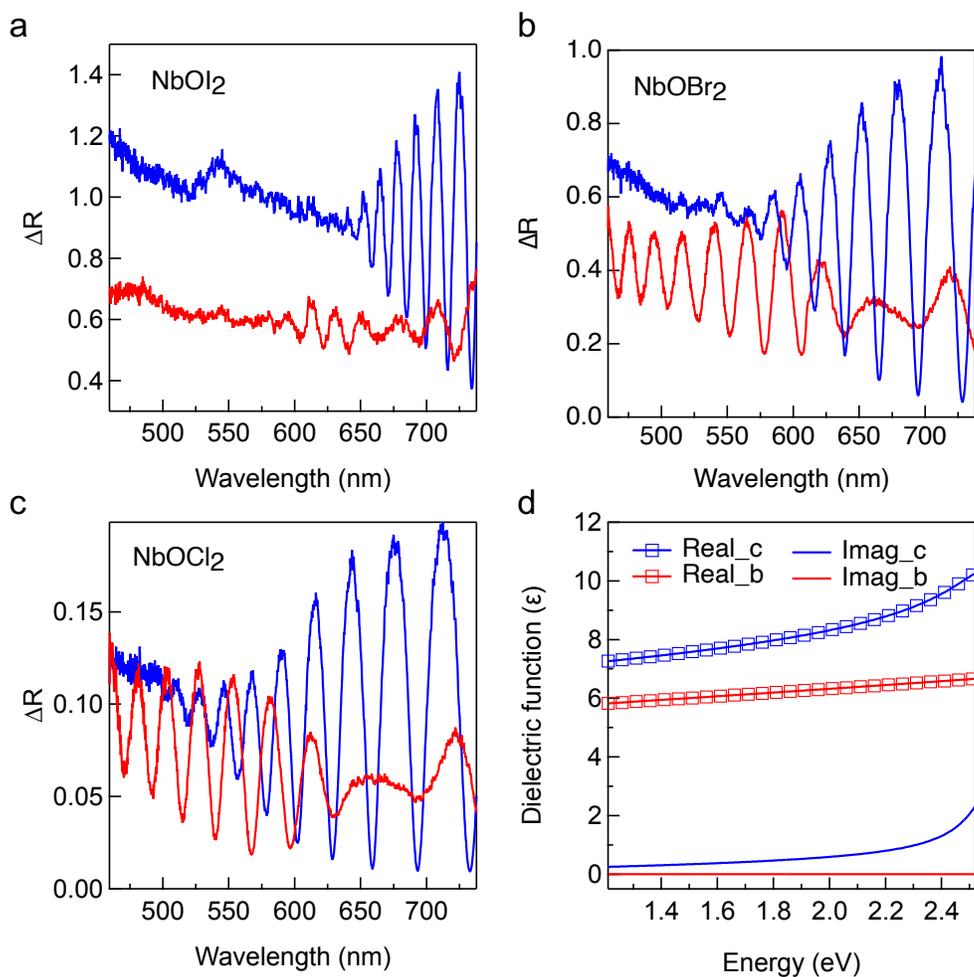

**Supplementary Fig. 14.** Differential reflectance (ΔR) of (**a**) NbOI$_2$, (**b**) NbOBr$_2$, and (**c**) NbOCl$_2$ THz-TDS sample. (**d**) Complex dielectric function of NbOBr$_2$. Signal from the polar axes is shown as a red curve, while the one from the non-polar axes is shown as a blue curve for each sample. To extract thickness information of NbOI$_2$ and NbOCl$_2$, we directly use their refractive index reported in ref. 4 and 5 respectively.